\documentclass[11pt]{article}

\usepackage[margin=1in]{geometry}
\usepackage{amsmath,amssymb}
\usepackage{graphicx}
\usepackage{booktabs}
\usepackage{natbib}
\usepackage{authblk}
\usepackage{setspace}
\usepackage{hyperref}
\usepackage{cleveref}
\usepackage{endnotes}

\title{The Geographic Moran Effect: Theory, Methods, and Interpretation}

\author[1]{Zhenzhi Jiao}
\author[1]{X. Angela Yao\textsuperscript{*}}
\author[2]{Jean-Claude Thill}

\affil[1]{Urban GeoAnalytics Lab, Department of Geography, University of Georgia, Athens, GA, USA}
\affil[2]{Department of Geography and Earth Sciences, University of North Carolina at Charlotte, Charlotte, NC, USA}
\affil[*]{Corresponding author: xyao@uga.edu}
\date{}

\begin{document}

\maketitle

\begin{abstract}
\noindent
The Moran effect traditionally refers to the synchronization of population dynamics driven by shared environmental forcing rather than direct interactions among populations. Although extensively studied in ecology, the Moran effect has received limited attention in geography, and its applicability has yet to be fully explored. In this study, we build upon the classical Moran effect in population ecology as a conceptual foundation, developing a more comprehensive conceptual framework in a geographic context. We introduce the new concept Geographic Moran Effect (GME), defined as temporally coherent and synchronized variation across geographic entities induced by common external forcing, with the magnitude and potentially the direction of local responses varying according to a spatially varying sensitivity field. A two-step spatiotemporal factor framework is developed to identify and interpret the GME. The framework first tests for the possible presence of GME-type synchrony in spatiotemporal data and then examines the GME and its spatiotemporal evolution by investigating its structure through spatial and temporal dimensions separately. A synthetic experiment validates the framework’s reliability, while an application to U.S. county-level presidential election data (2008–2024) demonstrates its ability to identify the GME and its spatial and temporal components. By redefining the Moran effect as a geographic theory and a distinct object of inference, this work provides a mechanism-oriented perspective on spatiotemporal analysis with broad applicability in geography, ecology, and public health, where explicitly identifying and interpreting GME-type synchrony is essential for understanding system-level dynamics.
\end{abstract}


\section{Introduction}
Spatial synchrony refers to the tendency of time-varying processes observed at different locations to fluctuate in concert, a phenomenon documented across ecological, environmental, and socio-economic spatiotemporal systems \citep{koenig2016temporally,liebhold2004spatial}. A foundational theoretical mechanism is the Moran effect, also known as Moran’s theorem, whereby spatially correlated or commonly experienced stochastic drivers synchronize otherwise independent local dynamics \citep{Engen2005,Hudson1999}. This idea traces back to early analyses of the Canadian lynx population cycles by \cite{elton1942ten} and was later formalized theoretically by \cite{moran1953statistical}, who demonstrated how correlated environmental stochasticity can synchronize spatially separated populations. Although originally developed in population ecology \citep{Blasius2000,Castorani2022}, the Moran effect provides a broader perspective on how geographical places may exhibit aligned temporal fluctuations in the absence of direct interaction or diffusion between them.

Conceptually, the Moran effect arises when a shared external temporal driver generates coherent variation across otherwise uncoupled local systems \citep{Goldwyn2011,Nicolau2022}. For instance, these shared temporal drivers may take the form of climatic fluctuations, economic shocks, public health events, political change, or other system-wide influences experienced across multiple locations. This perspective is particularly relevant to geography because spatially coordinated change is often observed across interconnected places, where similar temporal trajectories may otherwise be interpreted primarily through local interaction, diffusion, or spillover. This perspective implies that geographic co-fluctuation can emerge from common external forcing without requiring dependence generated through direct relationships among spatial entities.

Research on the Moran effect originated, and has been most extensively developed, in population ecology. It remains relatively underrecognized in other disciplines including geography, particularly in human geography \citep{walterGeographySpatialSynchrony2017}. One important reason is that the it has been primarily used to explain synchronized fluctuations among geographically separated populations, often observed across long distances or isolated systems such as islands \citep{Massie2015,Nicolau2022}. Its geographic extension raises two important considerations. First, neighboring entities may also exhibit synchronized variation because they are exposed to a shared external forcing. This is particularly important in geography because the Moran effect has traditionally been recognized in geographically separated or isolated systems, whereas synchrony among adjacent entities may be more readily attributed to relationships among neighboring units than to shared external forcing. A second challenge arises from geographic heterogeneity. The same forcing may be expressed differently across space, making externally driven synchrony less obvious than the conventional image of locations evolving together in the same way. For example, in political elections, the Moran effect in voting outcomes may arise from common national-level drivers, while regional contexts shape heterogeneous local responses \citep{Agnew1996,Li2021,Miller1994}. These two features can make Moran-type synchrony difficult to identify and interpret in geography. The critical geographic question is therefore whether observed co-fluctuation reflects common forcing expressed through geographically heterogeneous responses.

Accordingly, we define this geographically embedded form as the Geographic Moran Effect (GME). \textit{GME refers to temporally coherent variation across geographic entities induced by common external forcing, with the magnitude and potentially the direction of local responses varying according to a geographically structured sensitivity field}. This definition emphasizes the heterogeneous spatial expression of a shared temporal forcing across places and does not require direct interaction, diffusion, or propagation among spatial entities. It therefore provides a conceptual basis for distinguishing Moran-type synchrony from other forms of spatial synchrony.

Furthermore, although ecological studies have developed various approaches to analyze the Moran effect, they have predominantly focused on phenomenological synchrony\endnote{Phenomenological synchrony refers to the empirical observation or statistical detection of temporal co-fluctuation without separately identifying the mechanism that generates it.} rather than treating the Moran effect itself as a distinct object of inference. Consequently, these methods offer limited capacity to identify Moran-type synchrony when it occurs among neighboring entities and responses to the same external forcing vary geographically in magnitude or direction \citep{Ghislain2024,Hordley2025,Stenseth1999}. In other words, existing approaches are not designed to formally identify and test the GME mechanism, characterized by shared exogenous temporal forcing with spatially varying sensitivities. Their direct application to geographic systems is therefore limited because GME-type synchrony may be confounded with other forms of spatial or spatiotemporal dependence and heterogeneity. More specifically, the shared temporal component may be represented as latent common-factor dependence \citep{bai2009panel}, while geographically differentiated responses may be absorbed into unmeasured contextual effects associated with place \citep{fotheringhamMeasuringUnmeasurableModels2023a}.

To this end, we propose a two-step framework to identify and interpret GME. First, we develop a statistical testing procedure to assess whether GME-type synchrony is present in the data. Second, conditional on its presence, the framework estimates and interprets the GME through its common temporal factor and spatially varying sensitivity field. The rest of this paper is structured as follows: Section \ref{sec:background} discusses the theoretical and the methodological background; Section \ref{sec:method} introduces our methodology; Section \ref{sec:simulation} uses a synthetic dataset to validate the model’s effectiveness; Section \ref{sec:case} employs a case study of U.S. presidential elections to illustrate the model performance; Section \ref{sec:discussion} discusses the contribution and limitations of this study.

\section{Theoretical and methodological background}
\label{sec:background}

\subsection{Moran effect in geography}
The key phenomenon identified by Moran is that correlation in environmental forcing can induce correlation in population dynamics, implying that geographically isolated and dynamically uncoupled populations may still fluctuate in synchrony \citep{Koenig2002,Ranta1997}. This result is both counterintuitive and fundamental, as it shows that synchrony does not necessarily require biological connectivity or direct coupling mechanisms. Consider Moran’s original formulation for two uncoupled populations:
\begin{equation}
N_1(t+1)=f\bigl(N_1(t)\bigr)+\epsilon_1(t)
\label{eq:moran1}
\end{equation}

\begin{equation}
N_2(t+1)=f\bigl(N_2(t)\bigr)+\epsilon_2(t)
\label{eq:moran2}
\end{equation}
where $N_i(t)$ denotes the population size at location $i$; $f(\cdot)$ is the same renewal function for both populations.

Moran showed that when the environmental disturbances $\epsilon_1(t)$ and $\epsilon_2(t)$ are correlated, the resulting population fluctuations can inherit the same correlation:

\begin{equation}
\rho_{N_1,N_2} = \rho_{\epsilon_1,\epsilon_2},
\label{eq:moran3}
\end{equation}
where $\rho_{N_1,N_2}$ denotes the correlation between the two population dynamics and $\rho_{\epsilon_1,\epsilon_2}$ denotes the correlation between their environmental disturbances. Thus, the two populations can become synchronized through correlated environmental forcing without any explicit coupling between them. An important feature of this classical formulation is that the two populations are governed by the same renewal function, providing a theoretical benchmark in which synchrony arises from correlated environmental forcing despite the absence of direct coupling.

For geographic systems, we generalize this benchmark by allowing the response to the common forcing to vary across space. Building on the classical formulation of the Moran effect, the observed synchrony can be interpreted as the outcome of a shared environmental forcing acting across spatial entities, with location-specific responses governed by local system characteristics \citep{Bjrnstad1999,liebhold2004spatial,Ranta1997,Ripa2000}. This distinction separates the shared temporal forcing from the spatially varying response through which that forcing is geographically expressed. To express the conceptual decoupling of external shock-induced variation from local internal dynamics in a form amenable to statistical analysis, it is natural to represent this distinct spatiotemporal variation $u_{st}$ as:
\begin{equation}
u_{st} = a_s g_t,
\label{eq:gme}
\end{equation}
where $g_t$ denotes a common temporal factor capturing shared environmental fluctuations, and $a_s$ represents a spatially heterogeneous sensitivity field. Although this formulation is not explicitly stated in the original Moran literature, it provides a parsimonious representation consistent with its underlying mechanism. Similar separable structures have been widely adopted in ecological time-series analysis and spatial statistics \citep{Zuur2003,Bjrnstad1999,preisendorfer1988principal,Hannachi2007}. The loading $a_s$ is especially important for geographic systems because it allows a common forcing to be expressed differently across places. Consequently, GME does not require all locations to respond with the same magnitude or even the same direction. Locations are synchronized in the sense that their temporally coherent variation is linked to the same underlying temporal forcing, while geographically heterogeneous sensitivities determine how that forcing is locally expressed.

Classic ecological reviews distinguish three principal sources of synchrony: dispersal among populations, spatially correlated environmental forcing, and trophic interactions with mobile or spatially synchronized predators or competitors \citep{liebhold2004spatial,Ripa2000,Royama2005}. Methodologically, existing studies rely on three main approaches: detecting synchrony using correlation- and distance-based measures \citep{Bjrnstad1999,Koenig1999}; attributing synchrony by comparing population and environmental correlation structures \citep{Engen2005}; and assessing residual synchrony after accounting for environmental covariates \citep{Bjrnstad2001}. More recent work further incorporates phase-based and time-frequency analysis \citep{Cazelles2008,Sheppard2019}. Despite significant theoretical advances in identifying the spatial synchrony, much of the empirical literature remains centered on large-scale synchrony and the phenomenological detection of co-fluctuations \citep{Hordley2025,Larsen2025,Reuman2025}. In this context, phenomenological detection establishes the presence, strength, or temporal pattern of co-fluctuation, but does not by itself distinguish the latent common forcing from the spatially heterogeneous response field through which that forcing operates.

As it transpires, most existing studies do not treat the Moran effect as a standalone statistical component, but instead infer it indirectly by assessing whether observed population co-fluctuations track those of exogenous environmental variability. Moreover, methods developed in population ecology are not always directly transferable to geographic research, particularly in human geography, where spatial entities are strongly heterogeneous, neighboring units may simultaneously share external forcing and interact with one another, and responses to the same forcing may vary systematically in both magnitude and direction across space. This methodological gap raises a fundamental question: can GME be formally identified and represented as a distinct spatiotemporal component, rather than being indirectly inferred through post hoc comparisons between population and environmental synchrony? This question extends beyond methodological advancement and, more fundamentally, bears on our understanding of spatiotemporal processes, with implications for identifying causal effects. From a geographic perspective, the shared exposure of spatial entities to common external forcing implies that addressing this question becomes essential rather than optional. By treating GME as an identifiable structure, it becomes possible to reveal how shared temporal drivers are expressed across heterogeneous spatial contexts, uncovering systematic patterns of co-fluctuation that remain hidden under conventional approaches.

\subsection{Spatiotemporal dependence and common shocks}
Modeling spatiotemporal dependence is a central challenge in geographic data analysis, as observations are typically not independent across space and time \citep{Dub2013,Wikle2014}. Ignoring such dependence may lead to biased inference and unstable estimation. To address this issue, a large body of literature has developed graph-based models that incorporate neighborhood structures to represent spatial and temporal associations \citep{cressie2011statistics,Rue2005}. Among these, spatiotemporal Markov random fields (ST-MRF) provide a general probabilistic framework based on graph-structured dependence, while spatiotemporal conditional autoregressive (ST-CAR) models offer a widely used specification that combines spatiotemporal dependence with temporal autoregressive dynamics \citep{Lee2018}. These approaches are often implemented as structured random effects in hierarchical regression models.

Research on cross-sectional dependence and shared temporal variation has also been extensively developed in spatial econometrics \citep{arellano2003panel,Pesaran2006}. Traditional fixed-effects approaches, including time fixed effects and higher-order fixed effects, are widely used to control for shared temporal variation across spatial units. More flexible approaches, such as interactive fixed-effects models, latent factor models, and common correlated effects, further allow for heterogeneous responses and latent dependence structures in spatiotemporal data \citep{bai2009panel,moon2015linear}. More broadly, spatial econometric research has long distinguished among different forms of spatial dependence and spatial heterogeneity, while spatial panel models extend these considerations to repeated observations by jointly accommodating spatial and temporal structures. These approaches provide important tools for modeling cross-sectional dependence and shared temporal variation in large spatiotemporal systems. 

Notwithstanding their flexibility, spatiotemporal dependence models may reflect a mixture of underlying mechanisms, including spatial interaction, spillover, contextual background, and shared external influences, without explicitly distinguishing among them \citep{Darmofal2015}. This limitation becomes particularly important when the research objective is to identify a specific mechanism underlying observed dependence. Moreover, common temporal variation can be statistically represented or controlled through time effects and other common components without necessarily being assigned to a specific underlying mechanism \citep{angrist2009mostly,wooldridge2010econometric}. In practice, transient common shocks, broader shared trends, latent temporal structures, and externally forced synchrony may therefore enter similar statistical representations \citep{Chudik2015}. Although common-factor approaches account for shared temporal variation through latent factors, their primary inferential objective is generally the representation of cross-sectional dependence rather than the identification of a particular synchronization mechanism. Consequently, the presence of a common temporal component does not by itself establish that the observed co-fluctuation represents externally forced geographic synchrony.

These limitations become particularly important when the research objective is to distinguish externally forced synchrony from general spatiotemporal dependence or latent common dependence structures. As GME arises from shared temporal forcing across locations rather than solely from local associations among neighboring entities or latent common dependence structures embedded within generalized covariance representations, existing spatiotemporal dependence models and common-factor approaches may confound GME with other spatiotemporal processes. The methodological question addressed here is therefore narrower: whether externally forced synchrony can be explicitly identified and represented as a distinct component of observed spatiotemporal variation. This motivates the development of a modeling framework that explicitly identifies and represents GME as a distinct and interpretable component of spatiotemporal variation.

\section{Methodology}
\label{sec:method}

\subsection{Statistical testing of Geographic Moran effect}
\begin{figure}[htbp]
    \centering
    \includegraphics[width=\textwidth]{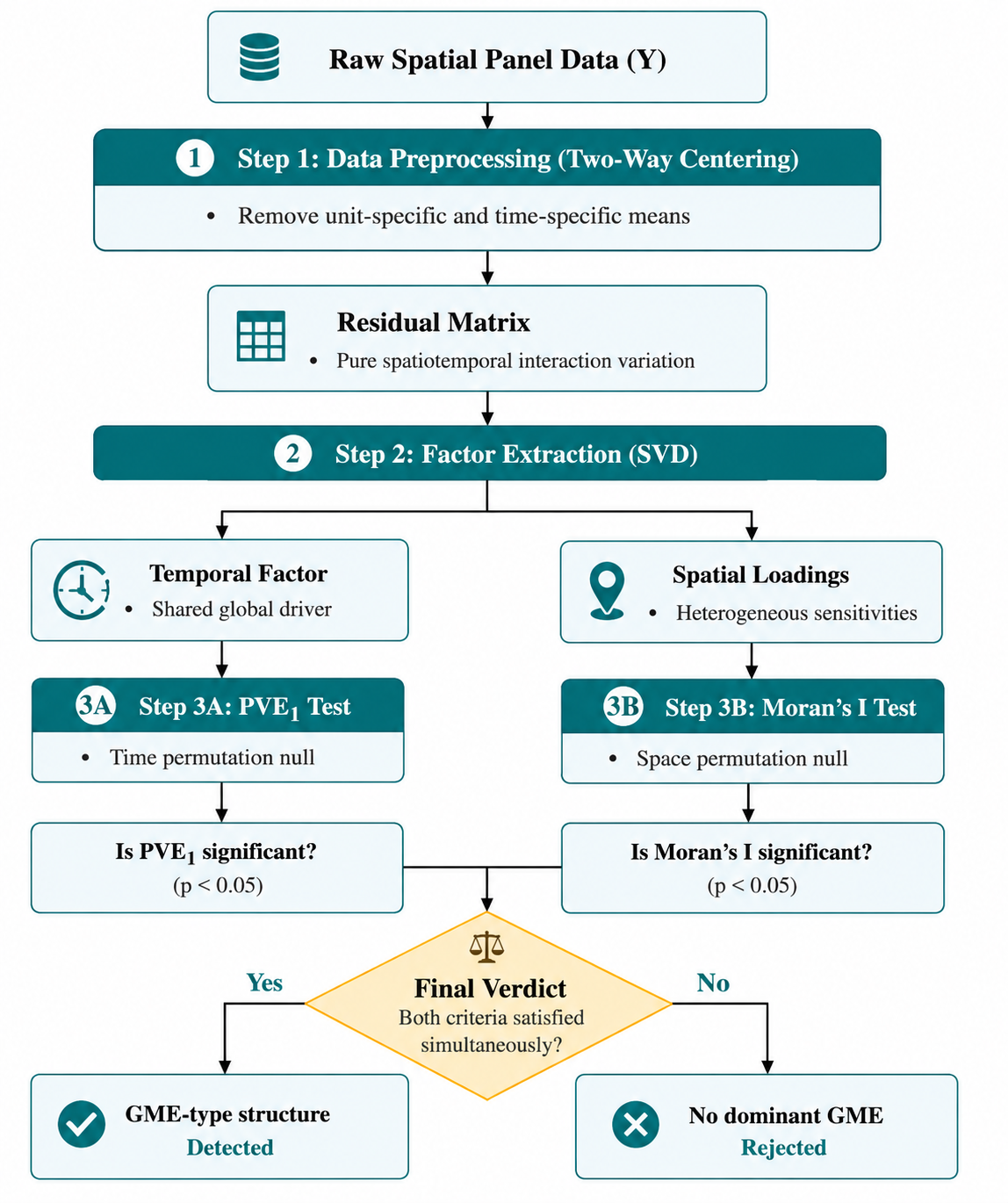}
    \caption{The workflow of step 1 in the two-step model}
    \label{fig:gme_workflow}
\end{figure}

In this study, we empirically test GME mechanism using a model-free procedure based on two criteria. As shown in Figure \ref{fig:gme_workflow}, first, two-way centering and Singular Value Decomposition (SVD) are used to examine whether the data exhibit a dominant interaction-type structure, as suggested by a relatively large leading singular component. Statistical support for this pattern is then evaluated through permutation tests of the variance explained by the leading component and through the spatial autocorrelation of its loadings. GME-type synchrony is regarded as supported only when both conditions are satisfied.

Let $y_{st}$ denote the observed response at spatial unit $s=1,\ldots,n$ and time $t=1,\ldots,T$, and define the data matrix $Y\in\mathbb{R}^{n\times T}$. To identify interaction-type variation, $Y$ is two-way centered by removing unit-specific and time-specific means. Let $R\in\mathbb{R}^{n\times T}$ denote the resulting residual matrix, with elements $r_{st}$. This operation eliminates purely spatial and purely temporal additive components, thereby focusing on cross-sectional--temporal interaction structures.

Under GME-type synchrony, spatiotemporal dependence arises from exposure to a shared temporal driver rather than from direct spatiotemporal interaction. Formally, the data-generating mechanism can be expressed as a common temporal forcing modulated by heterogeneous spatial sensitivities. This implies that the centered data matrix admits an approximate rank-one structure of the form

\begin{equation}
r_{st}=a_s g_t+\eta_{st},
\label{eq:gme_rankone}
\end{equation}
where $g_t$ represents a common temporal factor, capturing the direction and intensity of the common fluctuation shared across locations; $a_s$ represents spatially heterogeneous responsiveness, denoting differences in local sensitivity to this shared fluctuation and determining how each location amplifies or dampens the common signal; and $\eta_{st}$ captures idiosyncratic variation.

We evaluate this hypothesis using singular value decomposition (SVD) of the residual matrix $R$:

\begin{equation}
R=U\Sigma V^{\top},
\label{eq:svd}
\end{equation}
where $U$ contains orthonormal spatial singular vectors, $V$ contains orthonormal temporal singular vectors, and $\Sigma=\operatorname{diag}(\sigma_1,\sigma_2,\ldots)$ contains singular values ordered as $\sigma_1>\sigma_2>\cdots>0$. SVD provides the optimal low-rank approximation of $R$ in the Frobenius norm sense. In particular, the best rank-one approximation is given by

\begin{equation}
R^{(1)}=\sigma_1 u_1 v_1^{\top}.
\label{eq:rankone_svd}
\end{equation}
If GME-type synchrony is supported, the separable component $ag^{\top}$ aligns with this dominant singular component. Therefore, a large leading singular value indicates that the observed dynamics are primarily driven by a common temporal factor with spatially structured loadings.

The proportion of variance explained (PVE) by the first singular component is defined as

\begin{equation}
\mathrm{PVE}_1=
\frac{\sigma_1^2}{\sum_k \sigma_k^2}.
\label{eq:pve}
\end{equation}
Because $\sum_k \sigma_k^2=\lVert R\rVert_F^2$, this quantity measures the fraction of total residual variation attributable to the dominant separable structure. A large value of $\mathrm{PVE}_1$ provides stronger evidence for the presence of a GME-type structure, suggesting that residual dynamics are dominated by a common temporal factor rather than by independent noise or propagation-based interactions.

We use a permutation test to assess statistical significance. With this strategy, we construct a null distribution under the hypothesis of no cross-unit temporal synchrony. Time indices are permuted independently within each spatial unit:

\begin{equation}
r_{st}\longrightarrow r_{s\pi_s(t)},
\label{eq:permutation}
\end{equation}
where $\pi_s(\cdot)$ denotes a random permutation for unit $s$. This procedure preserves each unit's marginal temporal distribution and spatial variance while destroying the alignment of temporal shocks across entities. Under this null, any large singular value arises from random coincidence rather than systematic synchrony. The empirical distribution of $\mathrm{PVE}_1$ under this null is used to compute a one-sided $p$-value. A statistically significant and large $\mathrm{PVE}_1$ indicates the presence of a dominant shared temporal component. Specifically, we repeatedly permute the time indices within each unit, recompute $\mathrm{PVE}_1$ for each permuted sample, and compare the observed value with this empirical null distribution. The $p$-value is then given by the proportion of permuted samples that yield a $\mathrm{PVE}_1$ at least as large as the observed one.

Finally, to ensure that the detected synchrony is spatially structured consistently with an interaction-type mechanism, we examine the spatial loading vector $u_1$ using Moran's $I$. Statistical significance is assessed via a permutation test, in which the elements of $u_1$ are randomly permuted to generate an empirical null distribution. A significantly positive Moran's $I$ indicates that nearby locations exhibit similar responses to the common temporal factor, which is a defining feature of GME-type synchrony. We conclude that the data exhibit GME-type synchrony only when both conditions are satisfied, namely the presence of a dominant common temporal factor and spatially autocorrelated loadings.

\subsection{Examination of the geographic Moran effect}
\begin{figure}[htbp]
    \centering
    \includegraphics[width=\textwidth]{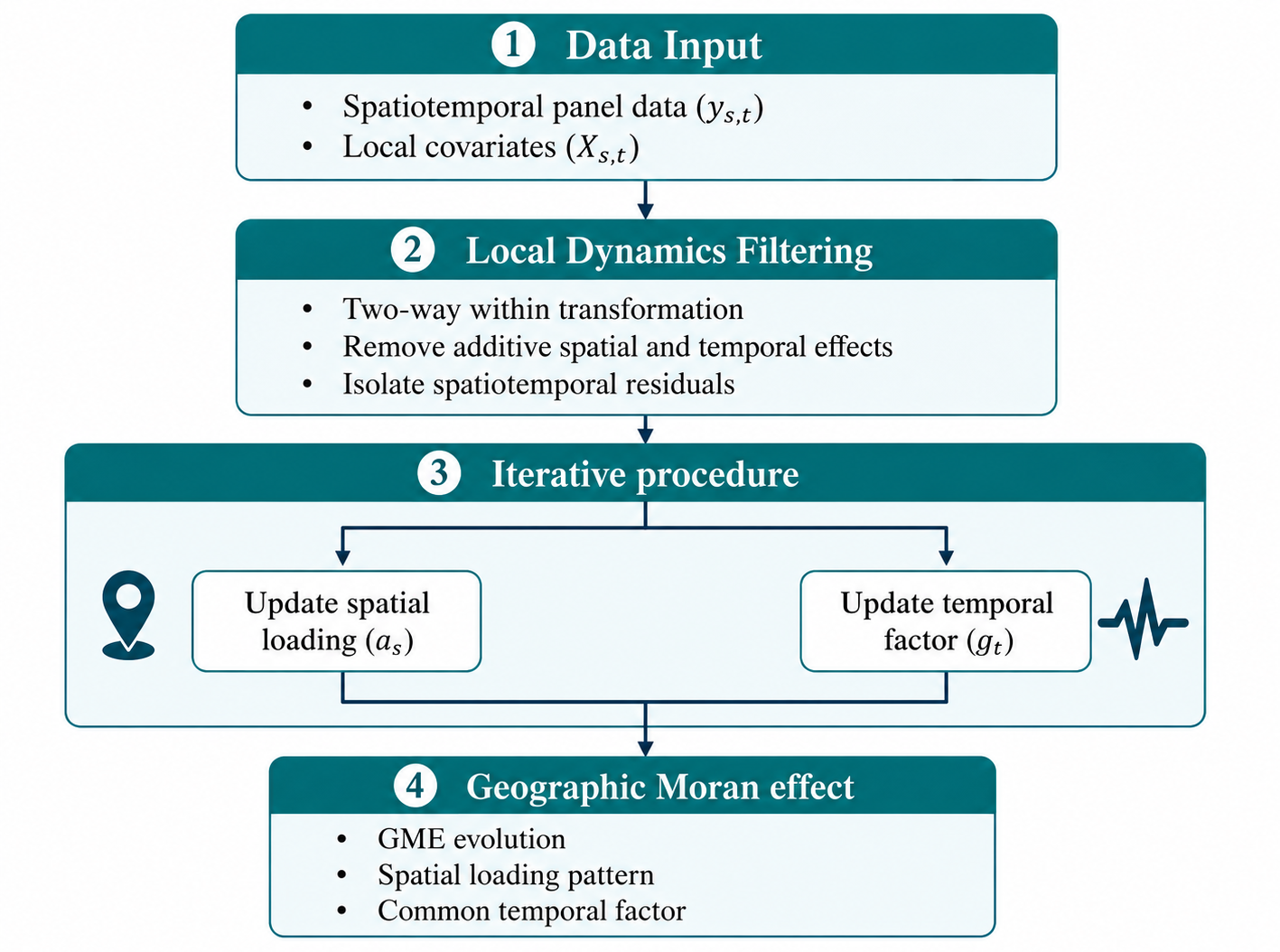}
    \caption{The workflow of step 2 in the two-step model.}
    \label{fig:f2}
\end{figure}

After establishing statistical support for GME-type synchrony, we aim to estimate GME itself and its spatiotemporal evolution, as well as recover its underlying structure in spatial and temporal dimensions separately (Figure \ref{fig:f2}). Figure \ref{fig:f2} summarizes the proposed estimation procedure. Starting from the observed spatiotemporal data and local covariates, we first apply a two-way within transformation to remove additive unit-specific and time-specific effects. We then remove variation attributable to the transformed local covariates to obtain partial residuals, and estimate the GME-type synchrony component through an iterative procedure that alternates between updating the spatial sensitivity field and the common temporal factor under smoothness constraints. The resulting decomposition yields GME together with its spatial and temporal components. 

Let $s=1,\ldots,n$ denote spatial entities and $t=1,\ldots,T$ index time points. The observed response is $y_{s,t}\in\mathbb{R}$, and $\mathbf{x}_{s,t}\in\mathbb{R}^p$ is a covariate vector. To maintain the same identification principle used in the first-stage GME test, the second-stage estimation begins with a two-way within transformation of both the outcome and each time-varying covariate:
\begin{equation}
\widetilde{y}_{s,t}=
y_{s,t}-
\bar{y}_{s,\cdot}-
\bar{y}_{\cdot,t}
+
\bar{y}_{\cdot,\cdot},
\label{eq:within_y}
\end{equation}
\begin{equation}
\widetilde{\mathbf{x}}_{s,t}=
\mathbf{x}_{s,t}-
\bar{\mathbf{x}}_{s,\cdot}-
\bar{\mathbf{x}}_{\cdot,t}
+
\bar{\mathbf{x}}_{\cdot,\cdot},
\label{eq:within_x}
\end{equation}
where $\widetilde{y}_{s,t}$ and $\widetilde{\mathbf{x}}_{s,t}$ denote the transformed outcome and covariates, respectively. This transformation removes additive spatial-unit effects and common temporal effects before the GME component is estimated. In particular, a temporal process that changes over time but affects all spatial units uniformly is removed as a time-specific main effect. The remaining variation therefore represents cross-sectional--temporal interaction after accounting for additive spatial and temporal components.

Using the transformed variables, we specify
\begin{equation}
\widetilde{y}_{s,t}=
\widetilde{\mathbf{x}}_{s,t}^{\top}\boldsymbol{\beta}
+
u_{s,t}
+
\varepsilon_{s,t},
\label{eq:gme_model}
\end{equation}
where $\boldsymbol{\beta}\in\mathbb{R}^p$ is the regression parameter, $u_{s,t}$ captures the GME-type synchrony, and $\varepsilon_{s,t}$ denotes an idiosyncratic error term with zero mean and finite variance.

To characterize the GME, we represent it as a separable rank-one spatiotemporal structure:
\begin{equation}
u_{s,t}=a_s g_t,
\label{eq:gme_component}
\end{equation}
where $a_s\in\mathbb{R}$ denotes the spatial loading at location $s$, and $g_t\in\mathbb{R}$ denotes the common temporal factor at time $t$.

Equivalently, in matrix form, let $\mathbf{U}\in\mathbb{R}^{n\times T}$ collect $u_{s,t}$, $\mathbf{a}=(a_1,\ldots,a_n)^{\top}\in\mathbb{R}^n$, and $\mathbf{g}=(g_1,\ldots,g_T)^{\top}\in\mathbb{R}^T$. The GME-type synchrony component can then be written as
\begin{equation}
\mathbf{U}=
\mathbf{a}\mathbf{g}^{\top}.
\label{eq:gme_matrix}
\end{equation}
The spatial loading vector $\mathbf{a}$ and the temporal factor $\mathbf{g}$ are estimated under spatial and temporal smoothness regularization, respectively, corresponding to a spatiotemporal Markov random field formulation. This representation provides a minimal parameterization of a shared temporal signal with spatially varying amplitude.

The factorization $a_s g_t$ is scale-indeterminate since, for any nonzero constant $c$, the transformation $(a_s,g_t)\mapsto(ca_s,g_t/c)$ yields the same product $u_{s,t}$, representing GME. Under the two-way formulation, an additional centering constraint is imposed on the spatial loading to separate geographically differentiated responses from the system-wide mean response. To ensure identifiability, the following constraints are imposed:
\begin{equation}
\frac{1}{n}\sum_{s=1}^{n}a_s=0,
\qquad
\frac{1}{T}\sum_{t=1}^{T}g_t=0,
\qquad
\frac{1}{T-1}\sum_{t=1}^{T}g_t^2=1.
\label{eq:identifiability}
\end{equation}
Under these constraints, the temporal factor $g_t$ is centered and standardized, so that it represents a zero-mean unit-variance common temporal signal. Consequently, the spatial loading $a_s$ directly quantifies the amplitude and direction of geographically differentiated temporally coherent variation relative to the system-wide average response. This normalization ensures that the magnitude of $a_s$ is uniquely defined and comparable across estimation runs. It also guarantees that variation in the loading field reflects genuine spatial heterogeneity in synchrony intensity rather than arbitrary rescaling of the temporal factor. Together with the two-way within transformation, these constraints ensure that purely additive temporal variation is not attributed to the GME component.

Let $\mathbf{W}\in\mathbb{R}^{n\times n}$ be a symmetric adjacency matrix and let
\begin{equation}
\mathbf{L}=
\mathbf{D}-
\mathbf{W}
\label{eq:laplacian}
\end{equation}
denote the graph Laplacian, where $\mathbf{D}=\operatorname{diag}(d_1,\ldots,d_n)$ with $d_i=\sum_j w_{ij}$.

Spatial regularity is imposed on $\mathbf{a}\in\mathbb{R}^n$ through the quadratic form
\begin{equation}
\mathbf{a}^{\top}\mathbf{L}\mathbf{a}=
\frac{1}{2}
\sum_{i=1}^{n}
\sum_{j=1}^{n}
w_{ij}(a_i-a_j)^2.
\label{eq:spatial_penalty}
\end{equation}
This penalty regularizes the loading field rather than the response itself. Its purpose is to encode smooth spatial heterogeneity of synchrony intensity while avoiding any recursive spatial dynamics imposed directly on $y_{s,t}$.

Temporal coherence is imposed on $\mathbf{g}\in\mathbb{R}^T$ via a first-difference precision penalty:
\begin{equation}
\mathbf{g}^{\top}\mathbf{Q}_t\mathbf{g},
\qquad
\mathbf{Q}_t=
\mathbf{D}_t^{\top}\mathbf{D}_t,
\label{eq:temporal_penalty}
\end{equation}
where $\mathbf{D}_t\in\mathbb{R}^{(T-1)\times T}$ is the first-order difference operator defined by
\begin{equation}
(\mathbf{D}_t\mathbf{g})_t=
g_{t+1}-g_t.
\label{eq:first_difference}
\end{equation}
This penalty is equivalent to imposing a Gaussian random-walk prior of order one on $\mathbf{g}$. It constrains the temporal factor to evolve gradually over time and penalizes abrupt inter-period variation. Consequently, the estimated factor captures persistent common temporal variation rather than short-scale oscillatory noise.

Parameters are estimated by minimizing a penalized least-squares criterion:
\begin{equation}
\mathcal{J}
\left(
\boldsymbol{\beta},
\mathbf{a},
\mathbf{g}
\right)
\sum_{s=1}^{n}
\sum_{t=1}^{T}
\left(
\widetilde{y}_{s,t}-
\widetilde{\mathbf{x}}_{s,t}^{\top}\boldsymbol{\beta}-
a_s g_t
\right)^2
+
\lambda_a\mathbf{a}^{\top}\mathbf{L}\mathbf{a}
+
\lambda_g\mathbf{g}^{\top}\mathbf{Q}_t\mathbf{g}
+
\gamma\lVert\boldsymbol{\beta}\rVert_2^2,
\label{eq:objective}
\end{equation}
where $\lambda_a>0$ controls spatial roughness of synchrony loadings, $\lambda_g>0$ controls temporal roughness of the common factor, and $\gamma>0$ is a ridge parameter stabilizing regression estimation under collinearity.

The objective function $\mathcal{J}$ is not jointly convex in $(\mathbf{a},\mathbf{g})$ due to the bilinear interaction. However, conditional on one block, the criterion is strictly convex in the other. This block-wise convexity enables an alternating minimization strategy. Let $\mathbf{R}\in\mathbb{R}^{n\times T}$ collect $R_{s,t}$, where we define partial residuals after removing the transformed covariate component as
\begin{equation}
R_{s,t}=
\widetilde{y}_{s,t}-
\widetilde{\mathbf{x}}_{s,t}^{\top}\boldsymbol{\beta}.
\label{eq:partial_residual}
\end{equation}

Conditional on $\mathbf{g}$, the spatial loading vector $\mathbf{a}$ is obtained by solving the linear system
\begin{equation}
\left(
\lVert\mathbf{g}\rVert_2^2\mathbf{I}_n
+
\lambda_a\mathbf{L}
\right)
\mathbf{a}=
\mathbf{R}\mathbf{g}.
\label{eq:update_a}
\end{equation}
The coefficient matrix is symmetric positive definite for $\lambda_a>0$, ensuring a unique solution. This step extracts the spatial amplitude field that best aligns the residual temporal variation with the current common temporal direction while respecting graph-induced smoothness. After each update, the spatial loading is centered to satisfy $\sum_s a_s=0$, ensuring that the estimated loading represents geographically differentiated sensitivity rather than an additive system-wide response.

Similarly, conditional on the spatial loading vector $\mathbf{a}$, the temporal factor $\mathbf{g}$ is obtained from
\begin{equation}
\left(
\lVert\mathbf{a}\rVert_2^2\mathbf{I}_T
+
\lambda_g\mathbf{Q}_t
\right)
\mathbf{g}=
\mathbf{R}^{\top}\mathbf{a}.
\label{eq:update_g}
\end{equation}
The coefficient matrix is symmetric positive definite for $\lambda_g>0$, yielding a unique solution. After each update, identifiability constraints are imposed by centering and rescaling $\mathbf{g}$, with corresponding adjustment of $\mathbf{a}$ to preserve the product $a_sg_t$. The spatial loading is subsequently re-centered to maintain the two-way identification constraint.

Given $\mathbf{a}$ and $\mathbf{g}$, define $u_{s,t}=a_sg_t$. The regression coefficient vector is obtained from the ridge-regularized least-squares problem
\begin{equation}
\widehat{\boldsymbol{\beta}}=
\arg\min_{\boldsymbol{\beta}}
\left\{
\sum_{s=1}^{n}
\sum_{t=1}^{T}
\left(
\widetilde{y}_{s,t}-
u_{s,t}-
\widetilde{\mathbf{x}}_{s,t}^{\top}\boldsymbol{\beta}
\right)^2
+
\gamma\lVert\boldsymbol{\beta}\rVert_2^2
\right\},
\label{eq:update_beta}
\end{equation}
which admits the following closed-form solution:
\begin{equation}
\widehat{\boldsymbol{\beta}}=
\left(
\widetilde{\mathbf{X}}^{\top}\widetilde{\mathbf{X}}
+
\gamma\mathbf{I}
\right)^{-1}
\widetilde{\mathbf{X}}^{\top}
\left(
\widetilde{\mathbf{y}}-
\mathbf{u}
\right),
\label{eq:beta_closedform}
\end{equation}
where $\widetilde{\mathbf{X}}$ and $\widetilde{\mathbf{y}}$ denote the stacked transformed covariates and outcome, respectively, and $\mathbf{u}$ is the estimated GME component.

The alternating minimization scheme produces a monotone decrease in the penalized objective function. Since the criterion is bounded below, the sequence of iterates converges to a stationary point. Although global optimality cannot be guaranteed due to non-convexity, empirical convergence is stable under moderate initialization conditions.

\subsection{Evaluation metrics}
Although the loading $a_s$ captures spatial sensitivity, $g_t$ describes temporal evolution, and $u_{s,t}$ reflects instantaneous effects, these quantities do not directly provide a stable, time-aggregated measure of the overall influence of GME at a given location. We therefore introduce two complementary summary measures that aggregate the GME-type synchrony component over time: the Moran Effect Index (MEI), which measures how much of a location's temporal variation is driven by the shared component, and the absolute effect (AE), which measures how large that shared effect is on average.

The spatial loading $a_s$ is the fundamental structural parameter of the GME-type synchrony component. Under the normalization constraint,

\begin{equation}
\operatorname{Var}_t(g_t)=1,
\label{eq:gt_variance}
\end{equation}
the temporal variance of the synchrony component at location $s$ satisfies

\begin{equation}
\operatorname{Var}_t(u_{s,t})
=
\operatorname{Var}_t(a_s g_t)
=
a_s^2\operatorname{Var}_t(g_t)
=
a_s^2.
\label{eq:ust_variance}
\end{equation}
Thus, $|a_s|$ represents the standard deviation of temporally coherent fluctuation at location $s$. The loading field therefore directly encodes spatial heterogeneity in synchrony intensity. Larger absolute values of $a_s$ indicate stronger susceptibility to the common temporal forcing. Because spatial regularization is imposed on $\mathbf{a}$, the estimated loading field reflects smooth spatial variation in synchrony magnitude rather than independent location-wise fluctuations.

To assess the relative dominance of GME-type synchrony in local temporal variation, we define the MEI as

\begin{equation}
\operatorname{MEI}(s)
=
\frac{\operatorname{Var}_t(a_s g_t)}
{\operatorname{Var}_t(y_{s,t})}.
\label{eq:mei}
\end{equation}
Using the variance identity above, this can be written equivalently as

\begin{equation}
\operatorname{MEI}(s)
=
\frac{a_s^2}
{\operatorname{Var}_t(y_{s,t})}.
\label{eq:mei_equivalent}
\end{equation}
MEI measures the proportion of total temporal variance at location $s$ attributable to the estimated synchrony component. Since both numerator and denominator are computed along the temporal dimension, the index is invariant to cross-sectional scale differences in outcome magnitude and enables comparison across heterogeneous spatial entities. When the synchrony component and residual variation are orthogonal in expectation, $\operatorname{MEI}(s)\in[0,1]$. Larger values indicate that common temporal forcing plays a dominant role in shaping the local temporal trajectory.

While MEI measures relative variance dominance, it does not preserve magnitude information on the original response scale. To quantify GME contribution in absolute terms, we define

\begin{equation}
\operatorname{AE}(s)
=
\frac{1}{T}
\sum_{t=1}^{T}
|u_{s,t}|
=
\frac{1}{T}
\sum_{t=1}^{T}
|a_s g_t|.
\label{eq:ae}
\end{equation}
where AE is the measure of the absolute effect. This quantity summarizes the average absolute contribution of GME to the observed trajectory at location $s$. Unlike MEI, it retains the physical entities of the response variable and therefore maintains substantive interpretability when outcome magnitudes carry practical meaning.

\section{Model validation with synthetic data}
\label{sec:simulation}

\subsection{Synthetic data design}
To evaluate whether temporally coherent spatial synchrony can be structurally separated from other forms of spatiotemporal dependence, we construct a controlled spatiotemporal data-generating process that embeds multiple spatial mechanisms simultaneously. The synthetic design represents three conceptually distinct components: (i) a separable common-shock synchrony structure (GME), (ii) spatial propagation through inter-entity interactions (interaction-type synchrony), and (iii) spatially autocorrelated background disturbances. In addition, to examine whether a common temporal covariate that is functionally related to its local counterpart affects recovery of GME, we introduce a global--local covariate structure in which a national-level temporal signal is embedded within a spatially varying local covariate. This layered design allows us to assess whether the proposed factor model can isolate multiplicative synchrony even when other spatiotemporal dependence and correlated global--local variation coexist.

Let $s=1,\ldots,n$ index spatial entities arranged on a regular two-dimensional lattice, and $t=1,\ldots,T$ index time. We first generate a global temporal covariate from an autoregressive process,

\begin{equation}
\bar{x}_t
=
\sum_{k=1}^{p_x}
\phi_k \bar{x}_{t-k}
+
\zeta_t,
\label{eq:global_covariate}
\end{equation}
where $\zeta_t$ is Gaussian white noise. A spatially patterned local deviation $\boldsymbol{\nu}_t\in\mathbb{R}^p$ is then generated through

\begin{equation}
\boldsymbol{\nu}_t
=
\left(
\mathbf{I}
-
\lambda_{\nu}\mathbf{W}
\right)^{-1}
\boldsymbol{\omega}_t,
\label{eq:local_deviation}
\end{equation}
where $\boldsymbol{\omega}_t$ is Gaussian white noise and $\lambda_{\nu}$ controls the spatial dependence of the local deviation. The first local covariate is constructed as

\begin{equation}
\mathbf{x}_{1,t}
=\bar{x}_t\mathbf{1}
+
\boldsymbol{\nu}_t.
\label{eq:local_covariate}
\end{equation}
Thus, the observed local covariate contains both the common temporal signal and geographically patterned local departures from that signal.

The outcome variable is generated according to

\begin{equation}
\mathbf{y}_t
=
\left(
\mathbf{I}
-
\rho\mathbf{W}
\right)^{-1}
\left(
\mathbf{X}_t\boldsymbol{\beta}
+
\gamma\bar{x}_t\mathbf{1}
+
\mathbf{a}g_t
+
\boldsymbol{\delta}_t
+
\boldsymbol{\xi}_t
\right),
\label{eq:simulation_dgp}
\end{equation}
where $\mathbf{y}_t=(y_{1,t},\ldots,y_{n,t})^{\top}\in\mathbb{R}^n$ is the outcome vector, $\mathbf{X}_t\in\mathbb{R}^{n\times p}$ denotes the covariate matrix at time $t$ whose $s$-th row is $\mathbf{x}_{s,t}^{\top}$, and $\boldsymbol{\beta}$ denotes regression coefficients. $\gamma$ represents the effect of the common temporal covariate $\bar{x}_t$, which enters the true data-generating process but is deliberately excluded from the fitted two-step model. $\mathbf{I}$ is the $n\times n$ identity matrix, $\mathbf{W}$ is an $n\times n$ row-standardized rook contiguity matrix, $\rho$ is the scalar spatial autoregressive coefficient, $\boldsymbol{\delta}_t\in\mathbb{R}^n$ is a spatially correlated disturbance, and $\boldsymbol{\xi}_t\in\mathbb{R}^n$ is an idiosyncratic Gaussian noise vector.

The GME-type synchrony structure is specified as a multiplicative interaction between a common temporal factor and a spatially varying loading field,

\begin{equation}
\mathbf{u}_t
=
\mathbf{a}g_t.
\label{eq:simulation_gme_vector}
\end{equation}

Equivalently, at the entity level,

\begin{equation}
u_{s,t}
=
a_sg_t,
\label{eq:simulation_gme_entity}
\end{equation}
where the temporal factor $g_t$ follows a stationary autoregressive process of order one,

\begin{equation}
g_t
=
\phi g_{t-1}
+
\eta_t,
\label{eq:simulation_temporal_factor}
\end{equation}
where $\eta_t$ is Gaussian white noise. This specification induces smooth and persistent temporal fluctuations. Although such persistence may occasionally produce visually repeating patterns over short horizons, it does not impose any deterministic periodicity. After generation, the temporal series is centered and scaled to unit variance to ensure that spatial heterogeneity in synchrony intensity is entirely captured by the loading field $a_s$. The spatial loading $\mathbf{a}=(a_1,\ldots,a_n)^{\top}\in\mathbb{R}^n$ is generated as a smooth random field defined on a spatial adjacency graph constructed using rook contiguity. Let $\mathbf{L}$ denote the corresponding graph Laplacian. The loading vector is sampled from a Gaussian distribution whose precision matrix is proportional to $\mathbf{L}$, inducing spatial smoothness across neighboring entities. The resulting field is centered and standardized.

To represent unobserved contextual processes that generate spatial dependence without propagation, we generate a spatially autocorrelated disturbance term $\boldsymbol{\delta}_t$ as

\begin{equation}
\boldsymbol{\delta}_t
=
\left(
\mathbf{I}
-
\lambda\mathbf{W}
\right)^{-1}
\boldsymbol{\varepsilon}_t,
\label{eq:spatial_disturbance}
\end{equation}
where $\lambda$ controls the degree of spatial error correlation and $\boldsymbol{\varepsilon}_t\sim\mathcal{N}\left(\mathbf{0},\sigma_{\varepsilon}^{2}\mathbf{I}\right)$ is Gaussian white noise. The spatially patterned deviation $\boldsymbol{\nu}_t$ used to construct the local covariate follows an analogous spatial-error mechanism, but operates on the predictor side rather than as a disturbance of the outcome.

In fitting the proposed model, only the local covariate matrix $\mathbf{X}_t$ is supplied, whereas the global temporal covariate $\bar{x}_t$ is deliberately omitted. The outcome and time-varying local covariates are then subjected to the two-way within transformation before estimation. This design therefore directly evaluates whether the underlying GME can be recovered when a common temporal covariate affects the outcome and is simultaneously embedded within an observed local covariate.

\subsection{Model validation}
The pre-test results indicate a clear GME-type synchrony structure in the data. The first singular component explains 57.1\% of the total variation ($\mathrm{PVE}_1=0.571$, $p=0.002$), indicating that the data are largely governed by a dominant common temporal factor. Furthermore, the corresponding spatial loading exhibits strong positive spatial autocorrelation (Moran's $I=0.952$, $p=0.002$). These findings jointly confirm the presence of GME-type synchrony.

\begin{figure}[htbp]
    \centering
    \includegraphics[width=\textwidth]{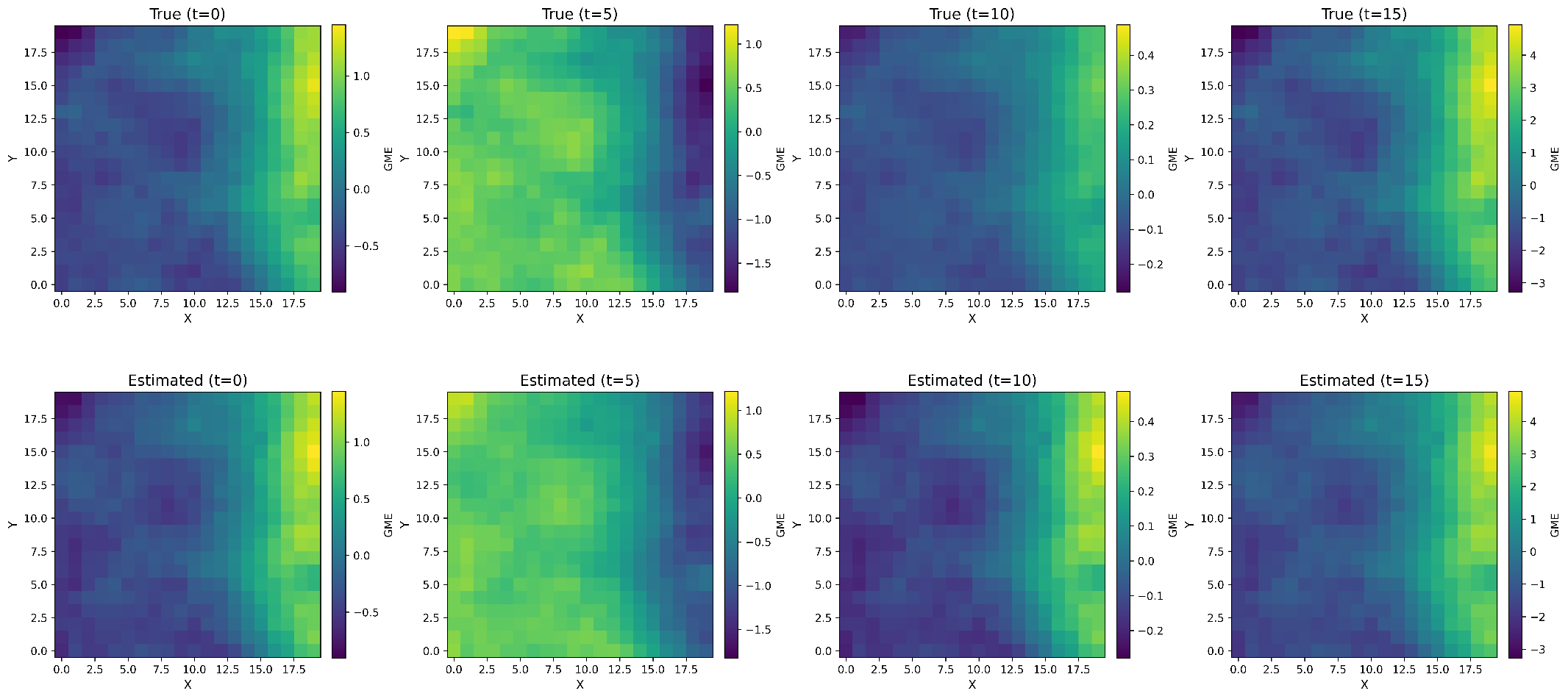}
    \caption{Recovery of the spatial loading field at different snapshots.}
    \label{fig:gme_evolution}
\end{figure}

We evaluate the ability of the proposed model to recover the GME-type synchrony structure by examining four representative snapshots ($t=0,5,10,$ and $15$). Each snapshot represents GME evaluated at a specific time point, and because it is estimated from the full time series rather than sequentially over time, the surface at $t=0$ is inferred from the overall temporal pattern rather than requiring earlier observations. Table~\ref{tab:gme_snapshots} reports the quantitative performance, showing consistently low RMSE and MAE and high $R^2$ across all time points. Figure~\ref{fig:gme_evolution} indicates the spatiotemporal evolution of GME and further illustrates that the estimated surfaces closely match the true GME, with the main spatial patterns and amplitude variations well preserved. Minor deviations are observed in localized regions, but the overall structure is accurately recovered. Importantly, the estimation remains stable across different temporal states, despite the presence of spatiotemporal interaction and spatially autocorrelated disturbances in the data-generating process. This indicates that the proposed framework is capable of isolating the GME component from other forms of synchrony.

\begin{figure}[htbp]
    \centering
    \includegraphics[width=0.9\textwidth]{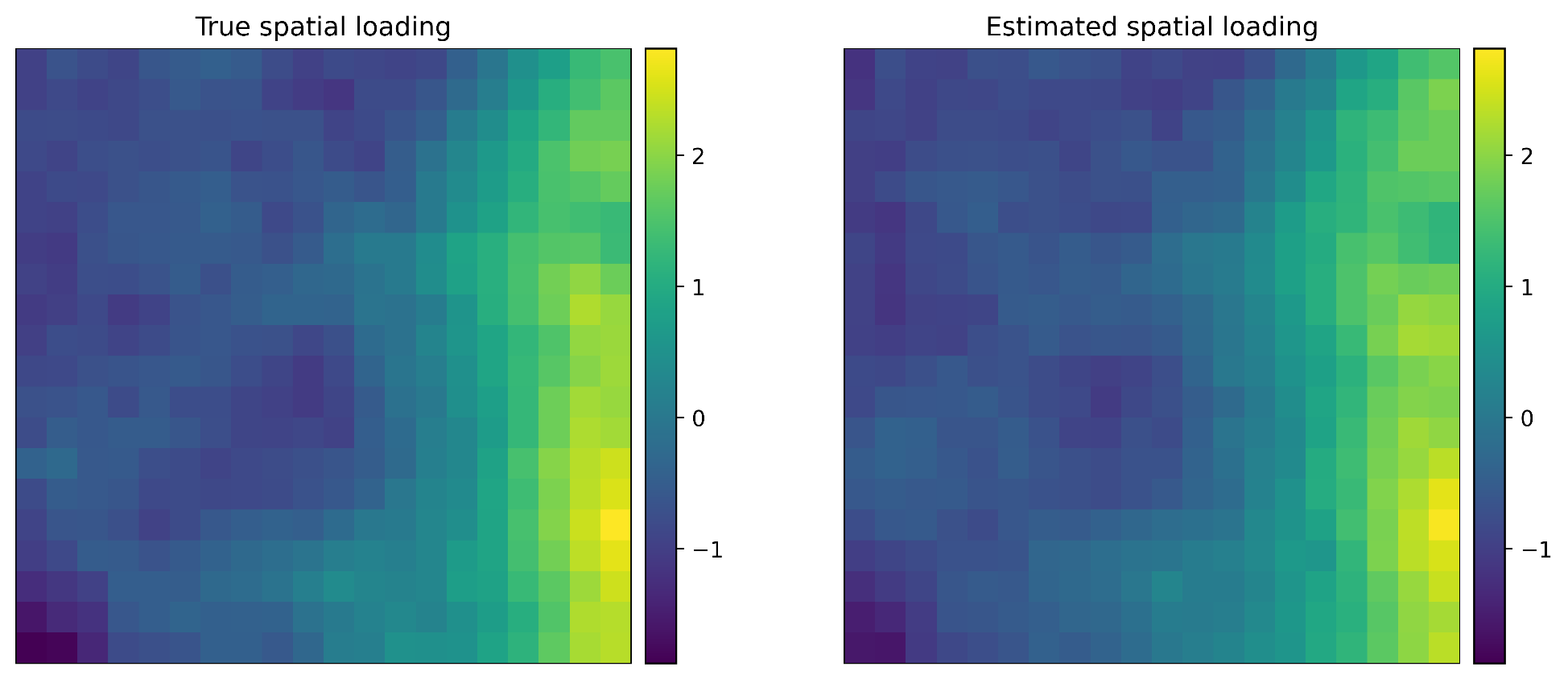}
    \caption{Recovery of the spatial loading field over the entire time span.}
    \label{fig:spatial_loading}
\end{figure}

\begin{figure}[htbp]
    \centering
    \includegraphics[width=0.9\textwidth]{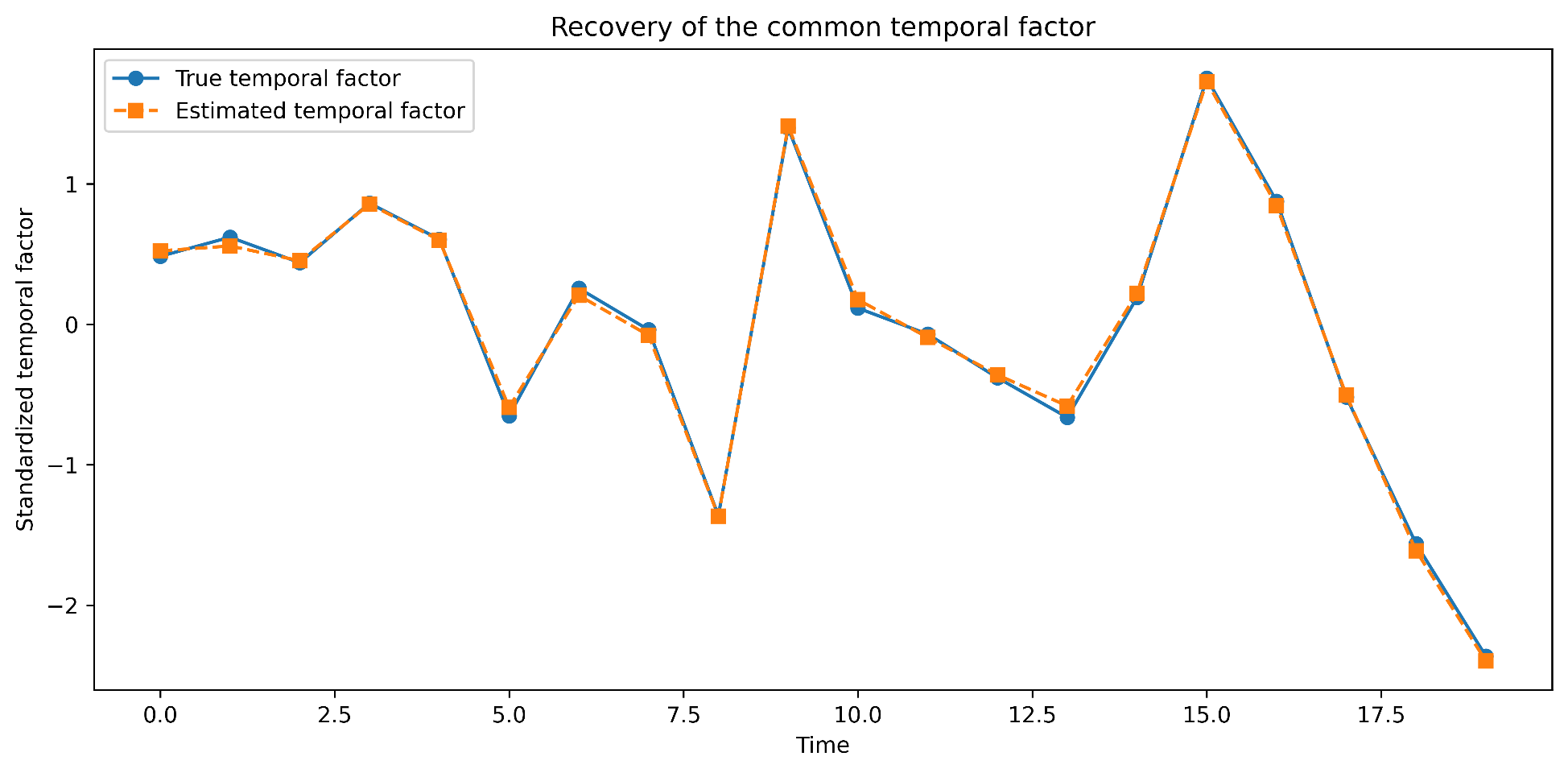}
    \caption{Recovery of the common temporal factor over time.}
    \label{fig:temporal_factor}
\end{figure}

\begin{figure}[htbp]
    \centering
    \includegraphics[width=0.9\textwidth]{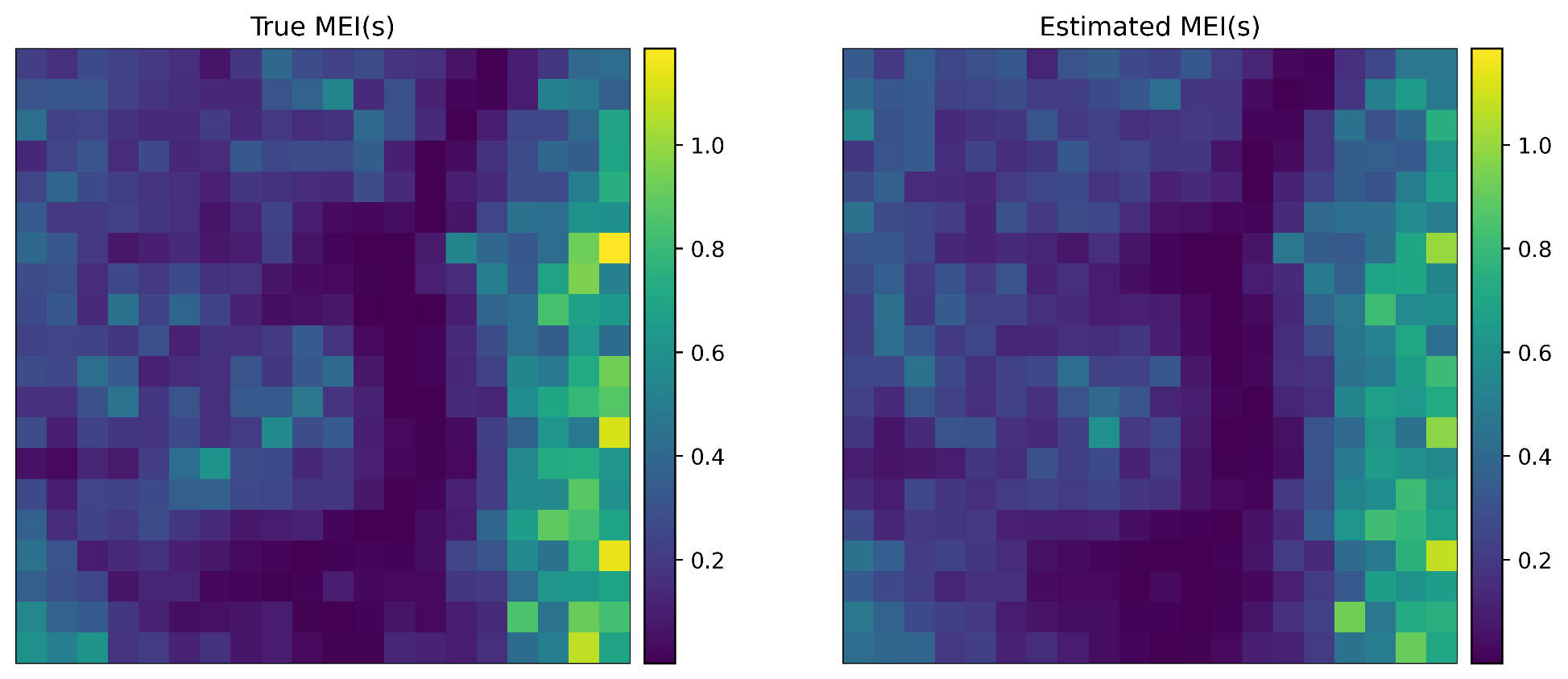}
    \caption{Estimated vs true local MEI across the entire time span.}
    \label{fig:mei}
\end{figure}

\begin{figure}[htbp]
    \centering
    \includegraphics[width=0.9\textwidth]{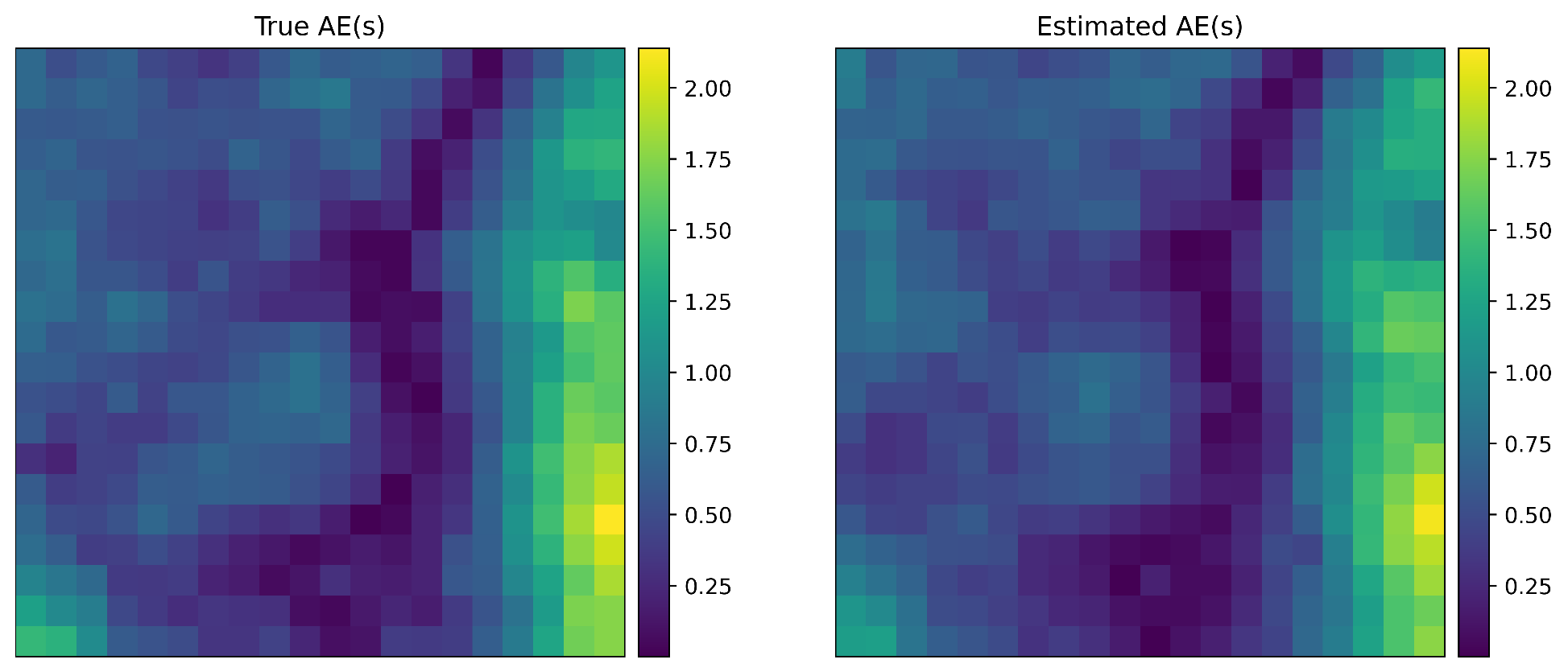}
    \caption{Recovery of the AE contribution across the entire time span.}
    \label{fig:ae}
\end{figure}

Table~\ref{tab:gme_components} reports the estimation accuracy for each component of the GME structure. Both the spatial loading and temporal factor are recovered with low error and high explanatory power, indicating accurate estimation of the underlying spatiotemporal components. As shown in Figures~\ref{fig:spatial_loading} and~\ref{fig:temporal_factor}, the estimated spatial loading reproduces the true spatial pattern, and the recovered temporal factor closely matches the true common trajectory. The full GME is also well reconstructed, with consistently low RMSE and MAE and high $R^2$. The derived measures, MEI and AE, are likewise estimated accurately across all evaluation metrics. Figures~\ref{fig:mei} and~\ref{fig:ae} provide visual confirmation: the estimated MEI reproduces the spatial distribution of relative synchrony importance, correctly identifying regions where temporally coherent variation dominates local dynamics. Similarly, the estimated AE closely matches the true spatial pattern and magnitude, preserving both large-scale gradients and localized variations in GME. Overall, these results show that the proposed approach reliably recovers GME, its spatial and temporal components, and their relative and absolute local manifestations under controlled conditions.

\begin{table}[htbp]
\centering
\caption{Estimations of the snapshots of GME.}
\label{tab:gme_snapshots}
\begin{tabular*}{0.75\textwidth}{@{\extracolsep{\fill}}lccc}
\toprule
Time ($t$) & RMSE & MAE & $R^2$ \\
\midrule
0  & 0.068 & 0.055 & 0.979 \\
5  & 0.099 & 0.078 & 0.976 \\
10 & 0.061 & 0.051 & 0.923 \\
15 & 0.212 & 0.169 & 0.985 \\
\bottomrule
\end{tabular*}
\end{table}

\begin{table}[htbp]
\centering
\caption{Recovery performance in the synthetic experiment over time.}
\label{tab:gme_components}
\begin{tabular*}{0.75\textwidth}{@{\extracolsep{\fill}}lccc}
\toprule
 & RMSE & MAE & $R^2$ \\
\midrule
Spatial loading $a_s$ & 0.120 & 0.095 & 0.985 \\
Temporal factor $g_t$ & 0.039 & 0.033 & 0.998 \\
GME $u_{s,t}$         & 0.123 & 0.083 & 0.983 \\
MEI                   & 0.072 & 0.052 & 0.897 \\
AE                    & 0.090 & 0.072 & 0.953 \\
\bottomrule
\end{tabular*}
\end{table}

\section{Case study}
\label{sec:case}

\subsection{Dataset and pre-processing}
\label{sec:dataset}

In this study, we conceptualize presidential elections as repeated national-level shocks conveying broad-based trends in the mindset of the U.S. population at large, and we focus on identifying GME in county-level Republican vote share, rather than modeling long-run temporal dynamics of electoral outcomes. We compile a dataset covering five U.S. presidential elections from 2008 to 2024 using publicly available election data hosted on GitHub.\endnote{\href{https://github.com/Josephjiao7/Geographic_moran_effect}{GitHub repository}} The dependent variable is the vote share received by the Republican (GOP) candidate, measured as the percentage of total votes cast at the county level in that year. Due to historical changes in administrative divisions and boundaries, we removed inconsistent geographic features from the dataset, resulting in a final sample of 3,099 counties across the U.S. mainland, which ensures geographic comparability across counties.

To control for potential confounding factors before estimating GME, we include several socioeconomic and demographic variables as covariates in the second stage of the model. Following the literature review, these covariates include median age, racial composition, educational attainment, median household income, poverty rate, homeownership rate, population density, and unemployment rate \citep{Amlani2021,Kahane2020,Li2021,Liu2020,Jiao2025}. All variables are obtained from the U.S. Census Bureau's American Community Survey (ACS) 5-year estimate datasets (2005--2009, 2008--2012, 2012--2016, 2015--2019, and 2020--2024), corresponding to the presidential election years analyzed in this study. To mitigate potential multicollinearity, variables with pairwise correlation coefficients greater than 0.7 are excluded based on correlation analysis. Prior to estimation, all variables are standardized to ensure comparability. The inclusion of these covariates is not essential to the identification of GME; rather, they are included here to demonstrate how the proposed framework can accommodate observed socioeconomic and demographic variation in an empirical application.

\subsection{Results and interpretation}
Evidence of GME-type synchrony is first examined in the empirical data. A dominant first singular component is observed, explaining 80.9\% of the total variation ($\mathrm{PVE}_1=0.809$, $p=0.002$). The associated spatial loading displays significant positive spatial autocorrelation (Moran's $I=0.605$, $p=0.002$), indicating that neighboring locations respond in a similar manner. These patterns collectively point to the presence of GME-type synchrony in the study area.

\begin{figure}[htbp]
    \centering
    \includegraphics[width=\textwidth]{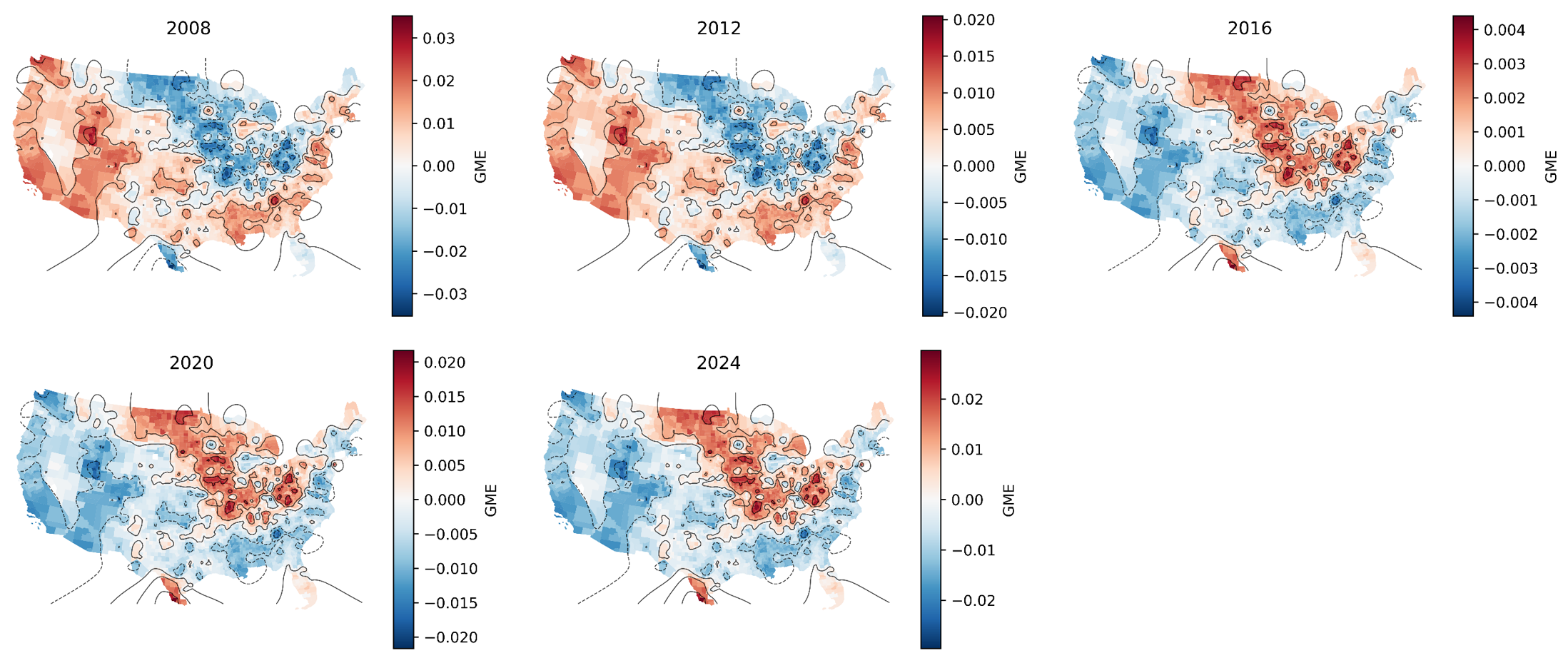}
    \caption{Spatiotemporal evolution and different snapshots of GME.}
    \label{fig:gme_empirical}
\end{figure}

Figure~\ref{fig:gme_empirical} visualizes the estimated GME based on county-level estimates, where spatial patterns are smoothed using radial basis function interpolation and highlighted with contour overlays. It illustrates the spatiotemporal evolution and the different snapshots of GME of county-level presidential election results. The estimated GME intensity captures whether county-level Republican vote shares move together under common national shocks or diverge across space. Positive values indicate synchronized shifts, whereas negative values reflect opposing regional responses. In 2008--2012, GME is highly heterogeneous, with broad positive and negative regions. This period corresponds to the political environment surrounding the election of President Barack Obama and the aftermath of the Global Financial Crisis. Despite strong national trends, many regions, including Appalachia, the Deep South, and parts of the Midwest, exhibited local resistance, with Republican support remaining stable or increasing. This reflects a regime in which local political and economic structures retained substantial autonomy and buffered common external shocks. A sharp structural break occurs in 2016, when GME weakens markedly and spatial coherence becomes minimal. This aligns with the electoral realignment associated with President Donald Trump, which reshaped traditional coalitions and redistributed support geographically. The near-zero GME suggests the absence of a dominant common force, with regions responding in highly idiosyncratic ways. In 2020--2024, GME becomes strongly positive and spatially uniform, indicating a high degree of synchronization across counties. This shift reflects the increasing ``nationalization'' of U.S. politics, driven by highly salient nationwide shocks such as the COVID-19 pandemic and intensified partisan polarization. Under this regime, counties across the country tend to respond in the same direction to common political signals, regardless of local context. Notably, although the Democratic candidate won the national popular vote in 2020, the unweighted county-level mean GOP vote share continued to increase. This pattern arises because GME captures synchronized temporal co-fluctuation across counties rather than population-weighted national vote totals. As each county is treated equally in the estimation, the resulting common temporal factor reflects broad inter-county voting dynamics instead of aggregate national electoral outcomes.

\begin{figure}[htbp]
    \centering
    \includegraphics[width=\textwidth]{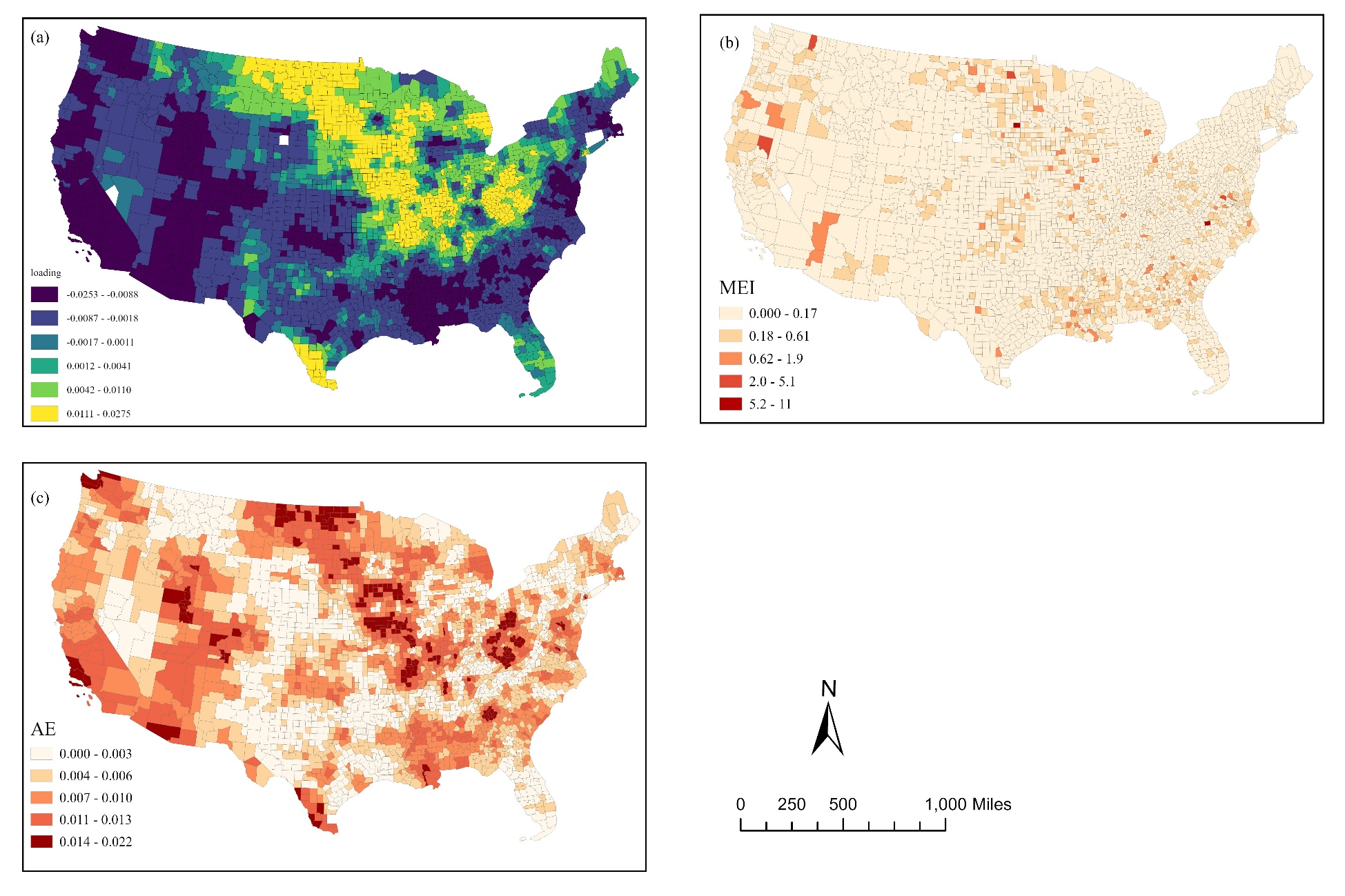}
    \caption{Results of the presidential election case study. (a) Estimated spatial loadings, representing local sensitivity to the common temporal factor. (b) MEI indicates the proportion of local temporal variation explained by GME. (c) AE captures the magnitude of GME-type fluctuations on the original scale.}
    \label{fig:gme_measures}
\end{figure}

Figure~\ref{fig:gme_measures}(a) presents the spatial distribution of the loading coefficient, capturing the geographically differentiated response of each county to the common temporal factor relative to the system-wide average response. High positive loadings are concentrated in the Midwest and Great Plains, indicating stronger-than-average responses in the same direction as the common temporal factor. In contrast, lower or negative loadings are more common in the West and parts of the Sun Belt, with negative values indicating relative responses in the opposite direction to the common temporal factor rather than simply weaker sensitivity. Overall, the pattern reveals clear spatial heterogeneity in how local entities translate common temporal shocks into observed outcomes. The MEI map in Figure~\ref{fig:gme_measures}(b) shows a spatially heterogeneous pattern, with most counties exhibiting relatively low values and higher values occurring in localized clusters across parts of the Southwest and Intermountain West, the Great Plains and Midwest, and scattered areas of the Southeast. Lower values are more widely distributed throughout much of the country, particularly across broad portions of the West and Northeast. Overall, high MEI values are relatively concentrated in specific inland areas rather than forming broad continuous regions. The AE map in Figure~\ref{fig:gme_measures}(c) displays a more spatially continuous pattern, with elevated values forming several broader regional concentrations across the northern and central Great Plains, the Upper Midwest, and parts of the Southeast and Appalachian region, together with additional localized concentrations in the Pacific Northwest and Southwest. Lower values are more common in parts of the interior West, portions of the Northeast, and several areas of the Southeast. Compared with MEI, AE therefore exhibits a substantially more spatially continuous geographic structure, indicating that the absolute magnitude of the estimated GME is regionally organized even where its relative contribution to local temporal variation remains modest.

From a policy perspective, the identified GME-type structure has important implications for understanding and responding to electoral dynamics. Because county-level voting behavior is partly driven by a shared temporal signal, shifts in national-level conditions such as economic performance, policy debates, cultural values, or major political events can induce coordinated changes across space. If future external shocks are favorable to Republican support, the model suggests that counties with higher positive loadings will exhibit stronger and more synchronized increases in Republican vote share, while areas with weaker or negative sensitivities will respond more modestly or diverge. Importantly, the largest marginal responses occur in counties with intermediate loadings. These regions are neither fully aligned with national trends nor entirely decoupled from them, making them particularly responsive to shifts in the common temporal signal. Thus, such areas may represent key targets for electoral mobilization, as relatively small changes in external conditions or campaign messaging can translate into disproportionate changes in vote share.\endnote{The interpretations are offered to illustrate potential implications of the findings and should not be construed as reflecting the authors’ political views.}

\begin{figure}[htbp]
    \centering
    \includegraphics[width=0.9\textwidth]{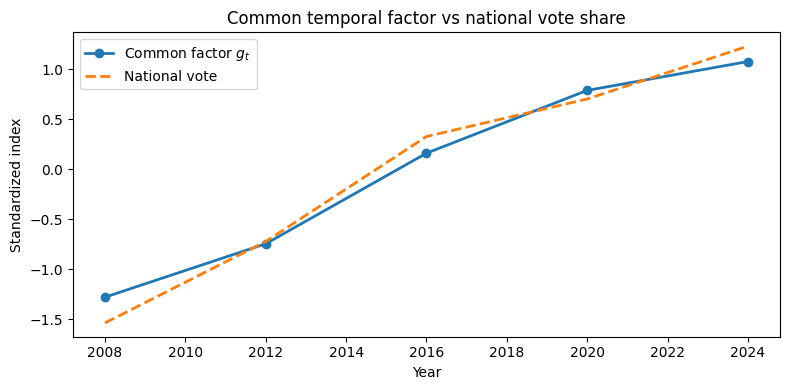}
    \caption{Comparison between the estimated common temporal factor and the national Republican vote share (both standardized).}
    \label{fig:common_factor_validation}
\end{figure}

To further assess the reliability of the model, we compare the estimated common factor $g_t$ with the national-level Republican vote share across presidential elections. A close correspondence between the two series provides additional evidence for the reliability of the estimated common factor. Because the two series are measured on different scales, both are standardized to have zero mean and unit variance before comparison. Figure~\ref{fig:common_factor_validation} shows that the estimated factor $g_t$ closely follows the trajectory of the national Republican vote share over time. In particular, both series exhibit similar temporal fluctuations across election years, including the decline around 2008--2012 and the subsequent increase after 2016. This strong temporal correspondence suggests that the extracted factor $g_t$ effectively captures the national-level political signal underlying county-level Republican vote share. In other words, the common factor identified by the model corresponds closely to nationwide partisan shifts in Republican electoral support. This result provides empirical support for interpreting $g_t$ as the shared temporal driver that generates GME-type synchrony across counties.

\section{Discussions and Conclusions}
\label{sec:discussion}

This study proposes the concept of GME and develops a corresponding spatiotemporal analysis framework for the detection and interpretation of this effect. First, it provides a formal test for the presence of GME in spatiotemporal data. Second, we estimate the spatiotemporal evolution and the overall GME, and interpret its spatial sensitivity field and common temporal factor. Using synthetic data, we demonstrate the effectiveness of the proposed model in recovering GME and its underlying spatiotemporal components, while distinguishing it from other mechanisms of spatial and spatiotemporal dependence. We further validate the framework through an empirical study of U.S. presidential elections from 2008 to 2024. The estimated GME captures a substantial component of the spatiotemporally coordinated variation in county-level Republican vote shares. The estimated spatial loadings for Republican support indicate that GME is particularly pronounced in the Midwest and parts of the Great Plains, largely overlapping with the broader Rust Belt and interior rural regions. Moreover, a comparison between the trajectory of the national Republican vote share and the estimated common temporal factor reveals a striking degree of alignment, providing further support for the validity of the proposed approach. These results suggest that the GME is not merely a statistical pattern of synchronized change, but a theoretically meaningful geographic process through which shared external forces generate spatially differentiated yet temporally coordinated dynamics.

The GME should be understood as one specific generating mechanism of spatial dependence, rather than as an alternative to existing spatial econometric models. Different forms of spatial dependence may arise from different underlying processes. Spatial autoregressive models in the outcome equation and related spillover models represent dependence transmitted through relationships among spatial units, whereas spatial error models capture spatial dependence through the disturbance structure without necessarily identifying its specific generating mechanism \citep{anselin1988spatial,lesage2009introduction}. GME represents a different source of spatial dependence: spatial units may co-fluctuate because they are exposed to a shared external temporal forcing, even without such inter-unit transmission. Spatial heterogeneity constitutes a second dimension of the GME. Geographically weighted approaches address a different question by characterizing spatial heterogeneity in relationships between observed covariates and outcomes \citep{Brunsdon1996,fotheringham2009geographically}. In the GME, the geographically differentiated response to the same external forcing is itself a form of spatial heterogeneity, represented by spatial variation in local sensitivities. However, this heterogeneity alone does not define the GME; the GME arises from its joint structure with the shared temporal forcing.

The temporal dimension alone is not what distinguishes GME, since spatial panel models already accommodate spatial and temporal dependence through spatial lags, temporal dynamics, fixed effects, spatially correlated disturbances, and related extensions \citep{elhorstAppliedSpatialEconometrics2010,elhorstSpatialEconometricsCrossSectional2014}. Similarly, the proposed formulation may resemble interactive fixed-effects or latent factor structures, but these approaches do not require the shared component to represent a specific external forcing process \citep{bai2009panel,Pesaran2006}. Their primary purpose is therefore to accommodate cross-sectional dependence and support statistical estimation, rather than to identify a particular synchronization mechanism. By contrast, GME treats externally forced synchronization itself as the object of inference: the common temporal factor represents shared forcing, while the spatial loading field represents geographically differentiated sensitivity to that forcing. This distinction is reflected in our synthetic experiment, where GME-type synchrony coexists with interaction-driven spatiotemporal dependence and spatiotemporally autocorrelated disturbances, allowing us to evaluate whether the GME component can be recovered when multiple mechanisms of spatiotemporal dependence are simultaneously present.

Although the Moran effect is a well-established mechanism in ecological studies and is commonly treated as a component of spatial synchrony, synchrony is typically analyzed from a mechanism attribution perspective, with different processes evaluated based on their explanatory power \citep{Nicolau2022,Sheppard2015}. The present study instead focuses on formally identifying and estimating GME-type synchrony as one specific spatiotemporal structure. More fundamentally, GME highlights two important features of geographic systems: common-forcing synchrony may occur among neighboring spatial entities, and the same forcing may generate geographically heterogeneous responses in magnitude and potentially in direction. From this perspective, the Moran-effect mechanism can be extended beyond its traditional ecological setting and interpreted as a form of structured geographic spatiotemporal variation that can be formally identified and modeled. The proposed framework therefore has potential relevance across a broad range of spatial systems in which shared external forcing and spatially differentiated responses are plausible. It provides a general approach for studying spatiotemporal synchrony phenomena driven by common external forcing in population dynamics, macroeconomic fluctuations, epidemic transmission, natural environment processes and collective social processes.

Beyond its immediate geographic application, the GME offers broader theoretical and methodological implications for the study of common-forcing synchrony across spatial systems. Theoretically, GME generalizes the classical Moran effect into a broader framework of common-forcing synchrony by allowing spatial entities to respond heterogeneously to the same external forcing in both magnitude and potentially direction. This extension is not restricted to geographic systems and can also provide a more general theoretical perspective for ecological and other spatially organized systems. Methodologically, the proposed framework provides a corresponding approach for identifying and estimating such synchronization through its shared temporal component and spatially differentiated sensitivities. Together, these theoretical and methodological extensions may be applicable to population dynamics, macroeconomic fluctuations, public health dynamics, environmental processes, and collective social dynamics. 

The current framework has two main limitations. First, it adopts a single-factor, rank-one representation of the GME component, in which geographically differentiated responses are linked to a single shared temporal factor. While this specification facilitates interpretation and aligns with the traditional Moran effect assumption, it may be overly restrictive in settings where multiple distinct temporal processes coexist, as these may be inadvertently compressed into a single latent component. Future research may extend the model by relaxing the rank-one constraint, for example by incorporating higher-rank structures or additional external information, to better disentangle multiple temporal drivers and improve the interpretability of GME. Second, the current framework is designed specifically to identify GME-type synchrony, even though other synchronization mechanisms may coexist in the observed system. Spatial synchrony in real-world systems may also arise from dispersal, diffusion, or direct interactions among entities. Future research could therefore extend the framework to jointly identify and characterize multiple synchronization mechanisms, providing a more comprehensive understanding of spatially coordinated dynamics.

\theendnotes

\section{Code and Dataset Availability}

The code and datasets used in this study are available at
\url{https://github.com/Josephjiao7/Geographic_moran_effect}.

\bibliographystyle{apalike}
\bibliography{references}

\end{document}